\documentstyle[aps,prl,preprint]{revtex}
\def\addcontentsline#1#2#3{\relax}

\newcommand{\beq}{\begin{equation}}
\newcommand{\eneq}{\end{equation}}
\newcommand{\bea}{\begin{eqnarray}}
\newcommand{\enea}{\end{eqnarray}}

\begin{document}
\title{Charge and current oscillations\\ in Fractional quantum Hall
systems with edges}
\author{J. Shiraishi$^{1}$\footnote{e-mail address:
shiraish@ginnan.issp.u-tokyo.ac.jp},
  \  Y.Avishai$^{1,2}$\footnote{e-mail address:
yshai@bgumail.bgu.ac.il}
  \  and M. Kohmoto$^{1}$\footnote{e-mail address:
kohmoto@issp.u-tokyo.ac.jp}\\
{\em
$^{1}$ Institute for Solid State Physics, University of Tokyo
Roppongi, Minato-ku, Tokyo 106 Japan\\
  $ ^{2}$ Physics Department,
   Ben Gurion University of the Negev, \\
   Beer Sheva, Israel}
}
\maketitle
\begin{center}
(\today)
\end{center}
\begin{abstract}
Stationary solutions of
the Chern-Simons effective field theory for the
fractional quantum Hall
systems with edges are presented for  Hall bar, disk and annulus.
In the infinitely long
Hall bar geometry (non compact case), the charge density
is shown to be monotonic inside the sample.
In sharp contrast, spatial oscillatory
modes of charge density are
found for the two circular geometries,
which indicate that in systems with compact geometry,
charge and current exist also far from the edges.

\end{abstract}
\newpage
\section{Introduction}
The present work focuses on charge and current distributions
in clean two dimensional electronic systems with edges
which are subject to strong perpendicular
magnetic fields.
Investigating the physics of the integer
or fractional quantum Hall effect
(QHE), and in particular, elucidation of
the precise charge and current profiles
in these systems is a fundamental problem
from both theoretical
and practical experimental points of view. The
quantum mechanical dynamics of electrons in two dimensional
systems at strong magnetic field is
characterized by two fundamental concepts.
The first one is the formation of Landau levels
which play an essential role in the studies of integer QHE.
The second, and probably more
profound is
the effect of the Coulomb interaction and
the emergence of the fractional QHE.

Indeed, a substantial theoretical effort has been
devoted to compute the charge and current distributions
within a realistic quantum mechanical picture.
For the integer QHE
in an infinitely long
Hall bar, a self consistent
formalism relating charge density
and electrostatic potential
has been suggested\cite{Macdonald}. Numerical solutions
of the pertinent equations
indicate that the charge density is monotonic decreasing inside
the Hall bar, and has a power law singularity ($|x-L|^{-1/2}$)
at the edges\cite{Macdonald}. In the
particular geometry of a semi-infinite plan, an analytic solution
using the Wiener-Hopf method is obtained\cite{Thouless}, and
the charge
density is found to have the same power
law singularity near the single edge.

For two dimensional electronic
systems at very strong magnetic fields the electron-electron
interaction plays a fundamental role, as it leads to the
fractional quantum Hall effect (and in fact, according to
the global picture\cite{Kivelson}, also to the
integer quantum Hall effect).
Hence, an evaluation of the charge and current profiles
in the fractional QHE
is somewhat more intricate.
One of the successful methods for describing
the fractional QHE
is the
effective Chern-Simons gauge field theory\cite{ChernSimons},
which includes the  Coulomb interaction
in a
non-perturbative manner. Employing this theory for
the calculations of charge and current distributions
is therefore distinct from various other
approximations used for treating these quantities
in interacting systems. These
include Hartree\cite{Wexler}, Hartree-Fock\cite{Halperin}
and others\cite{others}.

In the present work, calculation of charge and current profiles
are based on
the Chern-Simons field theory for treating the fractional quantum Hall
liquids. It turns out however, that
in order to preserve
gauge invariance, inclusion
of edges within this formalism must be done carefully.
One possible approach\cite{Wen} is to add a one dimensional
edge term to the original action. (See also \cite{org,han} and refs. therein.)
Once it is assumed that
all the important physical quantities are
concentrated near the edges, this
chiral edge action becomes a powerful tool for
studying the various edge effects. Thereby, the theory is
formulated through the Tomonaga-Luttinger liquid description,
whose experimental verification is under intense investigation
especially in tunneling experiments.
Recently, such tunneling effects were measured
and found to be consistent with theoretical predictions
\cite{FG}.

While studying the physics pertaining to
stationary quantum Hall states in which there
are `non-zero' currents flowing, one has no
apriori knowledge of whether or not
charge and current are concentrated near the edge.
In order to approach this problem,
we start our investigation from
the Chern-Simons effective theory which treats edge and bulk
properties on an equal footing\cite{KN}.
It leads, among other results, to a set of self-consistent
equations for charge and current distributions within
the two-dimensional sample.
In the most
general case, for filling factor $\nu=m/(mp+1)$ with integer
$m$ and even $p$,
there are $3m \times 3m$ coupled
integro-differential equations. Here we restrict ourself
to stationary solutions with $m=1$ (non-hierarchial case). For
clean systems with sufficiently high symmetry it ends up
with a single homogeneous eigenvalue problem determined by an
integro-differential equation.
This equation
 was derived in ref.\cite{KN} for the Hall bar geometry,
and appears
to be markedly different from its anlalogous one
for the integer QHE\cite{Macdonald}.

One of the interesting problems is
to study the spectrum of the eigenvalue equation thus obtained.
For the integer QHE, it is suggested that
the spectrum is continuous\cite{Macdonald}. On the other hand,
nothing is known for the fractional case so far.
If we assume that the spectrum is  continuous also
for the fractional case,
the
integro-differential equations can be solved
analytically order by order and it leads to a
charge distribution which is finite at the edges\cite{SK}.

We present below a general method for solving the
pertinent equations both for the integer\cite{Macdonald} and
the fractional\cite{KN} QHE systems.
Unlike the
previous order by order treatment, our method is
based on the Green function method for solving eigenvalue
problems.
The general method is then applied for some
specific systems of physical interest.

Our results indicate that for the infinitely long
Hall bar geometry, only a single eigenvalue exists,
which is physically acceptable,
and the corresponding charge distribution agrees with the
one obtained earlier using the analytic method\cite{SK}.
For the disk and annulus geometries, many physically
acceptable eigenvalues are present, and the corresponding
solutions behave as radial modes $\rho_{n}(r)$ with $n$
nodes between $R_{1}$ and $R_{2}$. This surprising appearance of
spatial charge oscillations in fractional quantum Hall
systems with compact geometries
has not been noticed before, and might shed a new
light on the physics of edge channels.

In Section 2, we explain the Green function method
which is used in solving the equations for the
charge density. For convenience, the method is
explained within the Hall bar geometry. The solutions
for the charge and current profiles in a
disk and an annulus, which comprise
the novel part of the present work are presented in Section 3.
A few specific topics are discussed in the Appendices.
In Appendix A, the Green function method for the integer QHE
system \cite{Macdonald} is discussed, while in Appendix B, the results
for the fractional QHE in the Hall bar geometry are
presented and shown to be consistent with those obtained
within an analytic treatment\cite{SK}. Finally, in Appendix C,
the Green function method is combined with a power expansion
technique which proves to be useful for the Hall bar geometry.

\section{The Green function Solution }

Consider an infinitely long Hall bar stretched along the $y$ axis between
the points $x=L_{1}$ and $x=L_{2}$ subject to a strong
magnetic field $H$ in the $z$ direction. In the stationary
QHE the charge density
$\rho$ depends only on $x$, and at present we are
interested in the density profile $\rho (x)$ for non-hierarchial
filling fraction $\nu=1/(p+1)$ with $p$ even.
Here $\rho$ is the difference between the charge distribution and its
average.
The equations derived in Ref.\cite{KN} couple the charge
density $\rho (x) $ and the current density $J_{y}(x)$, but if the
later is eliminated, a single integro-differential equation
for $\rho (x)$ remains, which reads
\bea                   \label{KNHB}
(8 \pi \nu c g^{-1})^{2}\frac{d^{2}} {dx^{2}} \rho(x)
 - \rho(x)=(4 \nu^{2} \xi g^{-1})
\frac{d^{2}} {dx^{2}} \int_{L{1}} ^{L_{2}} log|x-x'| \rho(x') dx'.
\enea
In this equation $c$ is the velocity of the
Chern-Simons gauge field and
$g^{-1}$ is the coupling constant of the Maxwell
term in the Chern-Simons Lagrangian. The constant
$\xi \geq 0$ comes from the solution of the
Poisson equation relating the electrostatic potential to
the charge density. Eq.(\ref{KNHB}) must also be
accompanied by the condition
\bea                 \label{Qzero}
\int_{L_{1}} ^{L_{2}} \rho(x) dx = 0.
\enea
It is useful to define $q \equiv (8 \pi \nu c g^{-1})^{-1}$ and
scale coordinates $x \rightarrow q x$ so that the integration limits
are $X_{i}=q L_{i}$. Eq.(\ref{KNHB}) then reads,
\bea                          \label{KNHB1}
\frac{d^{2}} {dx^{2}} \rho(x) - \rho(x) = \mu
\frac{d^{2}} {dx^{2}} \int_{X_{1}} ^{X_{2}} log|x-x'| \rho(x') dx',
\enea
where all the quantities appearing in the above equation are
dimensionless, including the constant $\mu \equiv \nu \xi/2 \pi c$.

Before proceeding with the solution of Eq.(\ref{KNHB1})
three general remarks are useful at this point.
1)
Eq.(\ref{KNHB1}) (and similar equations which are
obtained later for other geometries) is
a peculiar eigenvalue problem for the
charge density $\rho$, which can formally
be written as
\bea              \label{ev}
A \rho+\mu B \rho=0,
\enea
where $A$ and $B$ are certain linear operators and $\mu$ is
an eigenvalue. In Eq.(\ref{ev}), the operator $A$ is self
adjoint but $B$ is not. However, physics requires the existence
of real eigenvalues. We show below that Eq.(\ref{KNHB1})
can indeed be cast into an equation
having the same form as Eq.(\ref{ev}) in which $A$ is symmetric
and $B$ is symmetric and positive
definite. This problem has then only real
eigenvalues.
2)
The value of $\mu$
appearing therein cannot be chosen arbitrarily,
but must be selected from the
relevant set of eigenvalues.
On the other hand, following Eq.(\ref{KNHB1}),
the constant $\mu$
is composed of certain physical quantities such as mass, charge,
dielectric constant, filling factor {\em etc}.
If the spectrum is discrete, it might be tempting to formulate
here a kind of quantization rule. This is of course
too ambitious, since the basic theory is not an exact one.
On the other hand, the {\em sign} of $\mu$
should be consistent with its physical content.
3)
It is not useful to perform the second derivative
on the right hand side of Eq.(\ref{KNHB1}) and obtain
an integral equation. Indeed, the resulting
equation will have a very singular kernel which turn its
solution practically impossible.
Instead, we use the Green function
method which is so successful in solving eigenvalue problems
of the Sturm-Lioville type.
\bigskip

Without loss of generality we can set the Hall bar symmetrically
between the points $X_{1}=-qL$ and $X_{2}=qL$ (namely,
$L_{2}=-L_{1}=L$).
Furthermore, following
Refs. \cite{Macdonald,SK} we
limit our set of solutions to be
antisymmetric, $\rho (-x)=- \rho (x)$ which automatically
satisfy the
condition (\ref{Qzero}). The coordinate $x$ is then
limited within the interval $[0,q L]$ and the equation for the charge
density becomes,
\bea                      \label{KNHB2}
\frac{d^{2}} {dx^{2}} \rho(x) - \rho(x) = \mu
\frac{d^{2}} {dx^{2}} \int_{0} ^{qL}
\log \left| \frac {x-x'} {x+x'}\right| \rho(x') dx',
\enea
together with the condition $\rho (0)=0$. The value of
$\rho(qL)$ is not specified.

\noindent
We look for a (symmetric) Green function $G(x,x')=G(x',x)$ which,
formally can be considered as the inverse of the operator
$[\frac {d^{2}} {dx^{2}} - 1]$ with the appropriate boundary
conditions. Hence, it should satisfy
\bea        \label{GFEq}
&&\left[\frac {d^{2}} {dx^{2}} - 1\right]G(x,x')= \delta(x-x'),\\
&&G(0,x')=0.
\enea
Denoting respectively by $x_{<}$ and $x_{>}$ the smaller and larger
values of $x$ and $x'$, it is easily verified that
\bea      \label{GF}
G(x,x')=-\sinh(x_{<}) \cosh(x_{>})+ \alpha \sinh(x) \sinh(x'),
\enea
where the second term on the right hand side reflects the
freedom resulting from the
absence of a second boundary condition at $x=qL$.

We now express $\rho$ on the left hand side of Eq.(\ref{KNHB2})
in terms of $G(x,x')$. Formally we apply the operator
$G=[\frac {d^{2}} {dx^{2}} - 1]^{-1}$ on both side of (\ref{KNHB2}) and
use the formal identity
\bea                \label{GFI}
\left[\frac {d^{2}} {dx^{2}} - 1\right]^{-1} \frac {d^{2}} {dx^{2}}=
1+\left[\frac {d^{2}} {dx^{2}} - 1\right]^{-1}.
\enea
Strictly speaking, the operator $[\frac {d^{2}} {dx^{2}} - 1]$
has a zero eigenvalue corresponding to the function $\sinh x$ so that
its inverse must be defined within the subspace that is orthogonal
to this function. In so doing we might abandon some of the solutions
of the original equation (\ref{KNHB2}). We will see however that
the set of solutions is rich enough to capture the main physical
content.
With this point kept in mind
the integro-differential equation (\ref{KNHB2}) is transformed
into an integral equation,
\bea             \label{INTEQ}
\rho (x) = \mu \int_{0} ^{qL}  dx'\;
\log \left| \frac {x-x'} {x+x'}\right| \rho(x')
+ \mu
\int_{0} ^{qL} dx'' \int_{0} ^{qL} dx' \;
G(x,x'') \log \left|  \frac{x''-x'} {x''+x'}\right| \rho(x') .
\enea
Note that the right hand side of Eq.(\ref{INTEQ}) vanishes
at $x=0$ as required. The kernel of the integral equation does
not appear to be symmetric, so, apriori, there is no guarantee
that the eigenvalues $\mu$ are real. In order to proceed and
actually show it, it is convenient to write Eq.(\ref{INTEQ})
in its operator form,
\bea          \label{INTOPER}
\rho = \mu (1+G) L \rho ,
\enea
where the integral operators have their obvious coordinate
representations $\langle x|G|x'\rangle=G(x,x')$,
$\langle x|L|x'\rangle =
\log \left| \frac {x-x'} {x+x'}\right| $
(not to be confused with the system length) and $\langle x|1|x'\rangle
=\delta(x-x')$, all of them are symmetric.
It is not difficult to show that
the operator $-L$ is positive
definite, and therefore its square root $(-L)^{1/2}$ exists.
Eq.(\ref{INTOPER}) is then equivalent to the following
equation for $\eta \equiv (-L)^{1/2} \rho$,

\bea      \label{HBREAL}
\eta = \mu (-L)^{1/2} [-(1+G)] (-L)^{1/2} \eta,
\enea
whose kernel is symmetric. Only solutions corresponding to
positive eigenvalues $\mu$ are physically acceptable.
As we have commented after Eq.(\ref{GFI}), every solution
of Eq.(\ref{HBREAL}) is also a solution of Eq.(\ref{KNHB2}),
but the converse is not necessarily true. Numerical solutions
of Eq. (\ref{HBREAL}) will be discussed in appendix B.

\section{Oscillating charge and current distribution in disk and annulus}
We now use the Green function method introduced above to
solve the equations of Ref. \cite{KN} in a circular
geometry and demonstrate the occurrence of spatial charge
oscillations. The algorithm is demonstrated for a disk geometry,
and then, later on, some minor
modifications are introduced in order to study the annulus
geometry. We believe that for both systems
our results are experimentally relevant.

\noindent
Consider a clean disk of radius $R$ subject to a strong
perpendicular magnetic field such that the filling factor is
$\nu=1/(p+1)$ with $p$ an even number.
For systems with axial symmetry the charge density
$\rho$, the electrostatic potential $V$ and the
(tangential) current density $J_{\theta}$
depend only on the radial coordinate $r$. There is
of course no radial current. It is useful to define
the radial differential operator $D_{r}^{2} \equiv
\frac {d^{2}} {dr^{2}} + \frac {1} {r} \frac {d} {dr}$. Then,
we obtain the following set of equations,
\bea          \label{KNDISK}
&&\rho (r) - \frac {1} {q^{2}}  D_{r}^{2} \rho (r)
= \frac {4 \nu ^{2}} {g} D_{r}^{2} V(r) \\
&&V(r) = -\xi \int_{0}^{R} \phi (r,r') \rho(r') r' dr' \\
&&\phi (r,r') = \int_{0}^{2 \pi} \frac {d \theta} {\sqrt {r^{2}
+ r'^{2}-2 r r' \cos \theta}},
\enea
where $q \equiv (8 \pi \nu c g^{-1})^{-1}$. Of course,
the last two equations are just the solution of the Poisson equation
expressing the electrostatic potential $V(r)$ in terms of the
charge density $\rho (r)$ for systems with axial symmetry, with
$\xi \geq 0$. The charge density and the
 electrostatic potential are expected
to be finite in the disk.
In particular they should be regular at the origin. The weak
singularity of $\phi(r,r')$ at $r=r'$ is then integrable.

\noindent
Eliminating $V(r)$ from Eqs.(\ref{KNDISK}) and
transforming to dimensionless coordinates $x=qr$ we get a single
integro-differential equation for $\rho(x)$,
\bea           \label{KNDISK1}
[D_{x}^{2}-1]\rho(x)= \mu D_{x}^{2} \int_{0}^{qR}
\phi (x,x') \rho(x') x' dx',
\enea
where $\mu \equiv \nu \xi/2 \pi c>0$.

\noindent
We look for a (symmetric) Green function $G(x,x')=G(x',x)$ which,
formally can be considered as the inverse of the operator
$[D_{x}^{2}-1]$ with the appropriate boundary
conditions (regularity at $x=0$). Hence, it should satisfy
\bea        \label{GFCIRC}
&&[D_{x}^{2}-1]G(x,x')= \frac {\delta(x-x')} {x},\\
&& G(0,x'):\;{ \rm finite}.
\enea
Denoting respectively by $x_{<}$ and $x_{>}$ the smaller and larger
values of $x$ and $x'$, it is easily verified that
\bea      \label{GFDISK}
G(x,x')=-I_{0}(x_{<}) K_{0}(x_{>}) + \alpha I_{0}(x) I_{0}(x') ,
\enea
where $I_{0}$ and $K_{0}$ are the modified Bessel functions.
The second term on the right hand side reflects the
freedom resulting from the
absence of a second boundary condition at $x=qR$.

\noindent
We can now repeat the same procedure which led from Eq.(\ref{KNHB2})
to Eq.(\ref{INTEQ}), and later to Eq.(\ref{HBREAL}),
with some slight modification due to the
presence of the volume element $x\;dx$ in the relevant integrals.
Thus we define the operators $G$ and $\Phi$ such that their
configuration space representations are
$\langle x|G|x'\rangle =\sqrt x G(x,x') \sqrt{ x'}$ and
$\langle x|\Phi |x'\rangle
=\sqrt x \phi(x,x') \sqrt{ x'}$. It is not difficult to show that
the operator $\Phi$ is positive definite, so its square root
is well defined. The function $\eta \equiv
\Phi ^{1/2} x^{1/2} \rho $ then
satisfies the integral equation
\bea               \label{ETAEQ}
\eta = \mu \Phi^{1/2} (1+G) \Phi^{1/2} \eta,
\enea
whose kernel is symmetric.

\noindent
For an annulus with radii $R_{1}<R_{2}$ the Green function
has much more freedom because the origin is not reached.
Instead of (\ref{GFDISK}) we may now have
\bea      \label{GFANNULUS}
G(x,x')=-[\alpha I_{0}(x_{<}) K_{0}(x_{>}) +
\beta K_{0}(x_{<}) I_{0}(x_{>})]
+ \gamma I_{0}(x) I_{0}(x') + \delta K_{0}(x) K_{0}(x') ,
\enea
with $\alpha + \beta=1$.\\

\noindent
We have solved Eqs.(\ref{GFANNULUS}) numerically
for the disk and the annulus geometry using
$N=160$ Gaussian integration points for the radial integration.
As a credibility check of our numerical procedure we have
confirmed that
the numerical value of the positive eigenvalues is independent
of $N$ .

\noindent
The operator appearing on the right hand
side of equation (\ref{ETAEQ})
is real and symmetric. One can then
look upon the eigenvalue equation
as an equation for a rope which is not attached at its ends.
This analogy is somewhat expected, since equation
(\ref {KNDISK1}) is very similar to an inhomogeneous wave
equation for stationary solutions. The resulting solutions
$\rho_{n}(x)$ corresponding to positive eigenvalues
$\lambda_{n}=1/\mu_{n}$, $n=0,1,2,..$ with
$\lambda_{n}<\lambda_{n+1}$ appear to have $n$ radial nodes.
The lowest mode $n=0$ cannot satisfy the condition of zero
total charge (\ref{Qzero}), and hence
it is physically unacceptable. All the other modes indeed
have zero total charge, and can therefore be regarded as
representing the pertinent radial charge density.

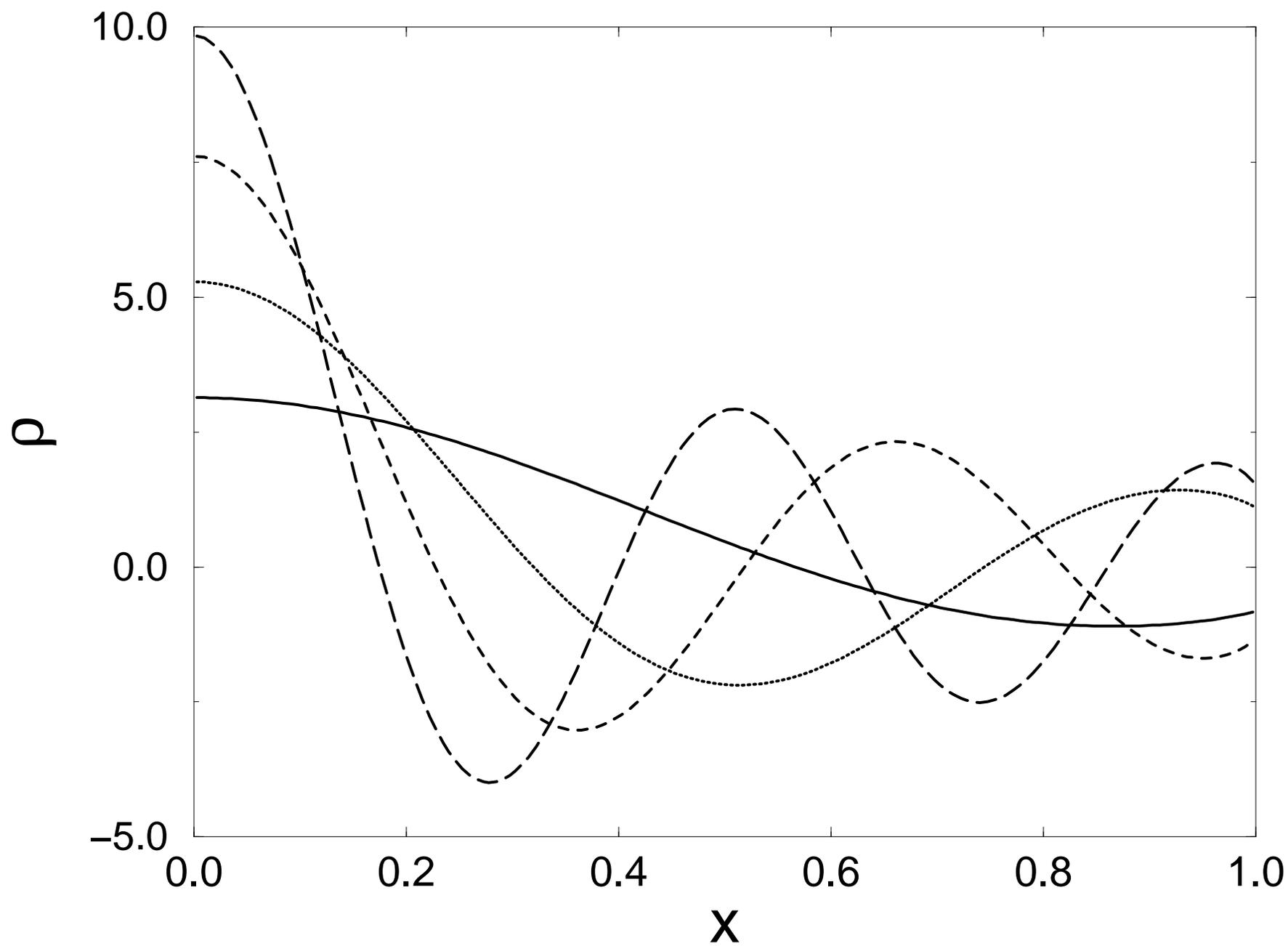
\begin{figure}[t]
\caption{Charge distribution in a disk. The
parameters are $qR=1$ and $\alpha=0$
(see equation (\ref{GFDISK})). The four modes
correspond to the four positive eigenvalues
$\lambda_{n} \equiv 1/\mu_{n}$
of equation (\ref{ETAEQ}), with $n=1,2,3,4$ which
also counts the number of radial nodes.\label{Fig1}}
\end{figure}

\noindent
For the disk
geometry we chose $qR=1$ in Eq.(\ref{KNDISK1}) and
$\alpha=0$ in the expression (\ref{GFDISK}) for the Green
function. The charge density profile for the lowest four
modes $n=1,2,3,4$ is displayed in Fig.~\ref{Fig1}. The basic
features of these solutions are as follows:
1) The charge density
is maximal at the center.
2)
 The mode $n=1$ with one zero reminds
us of the charge distribution in the Hall bar geometry although
it does not have a sharp peak at the edge.
3)
 Most notably, there
are numerous oscillatory solutions which,
to the best of our knowledge, were not predicted so far.

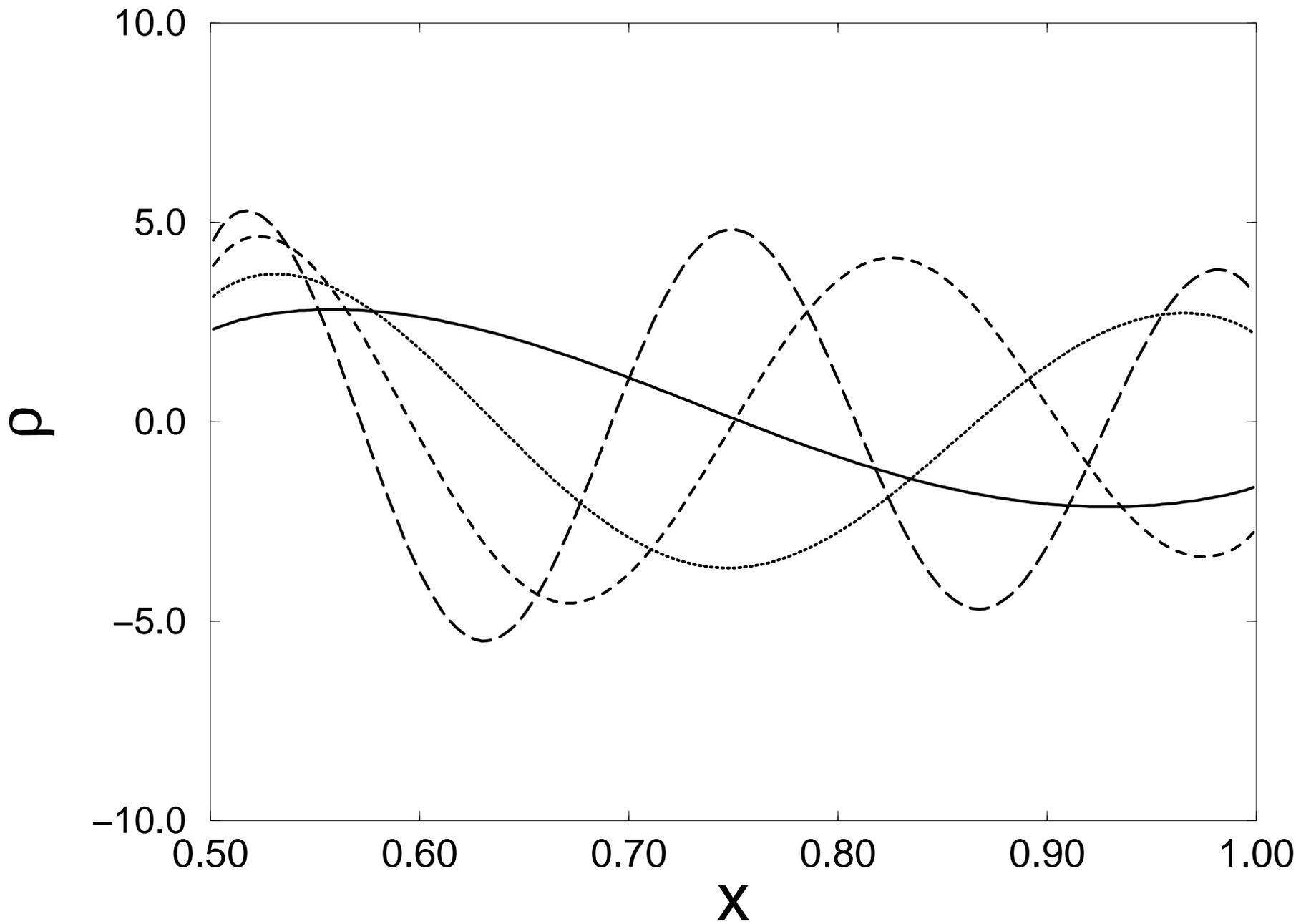
\begin{figure}[t]
\caption{
Charge distribution in an annulus\label{Fig2}. The
parameters are $qR_{1}=0.5$, $qR_{2}=1$ and $\alpha=1$,
$\beta=\gamma=\delta=0$ (see equation (\ref{GFANNULUS})).
The four modes are as in figure \ref{Fig1}.}
\end{figure}

\noindent
For the annulus
geometry we chose $qR_{1}=0.5$ and $qR_{2}=1.0$ as integration
limits instead of $0$ and $qR$
in Eq.(\ref{KNDISK1}). We also take
$\alpha=1, \beta=\gamma=\delta=0$
in the expression (\ref{GFANNULUS}) for the Green
function. The charge density profile for the lowest four
modes $n=1,2,3,4$ is displayed in Fig.~\ref{Fig2}. The basic
features of these solutions are different from the ones
found for the disk in that they do not have a maximum
at the inner edge. From that point of view, a disk
cannot be considered as an annulus with vanishingly small
inner radius. The large value of the charge at $R=0$ is
attributed to the weak (but integrable) singularity of
$K_{0}(x)$ near $x=0$. However, here again there
are numerous oscillatory solutions. We know \cite{SK}
and shall also see in Appendices B and C
that in the Hall bar geometry, there is only a single
solution, which is monotonic.  The novel result
of spatial oscillatory charge density must
then be related to the fact that the pertinent physical systems
are compact. It then belongs to the realm of mesoscopic and
quantum dot physics, which, nowadays,
is experimentally accessible.

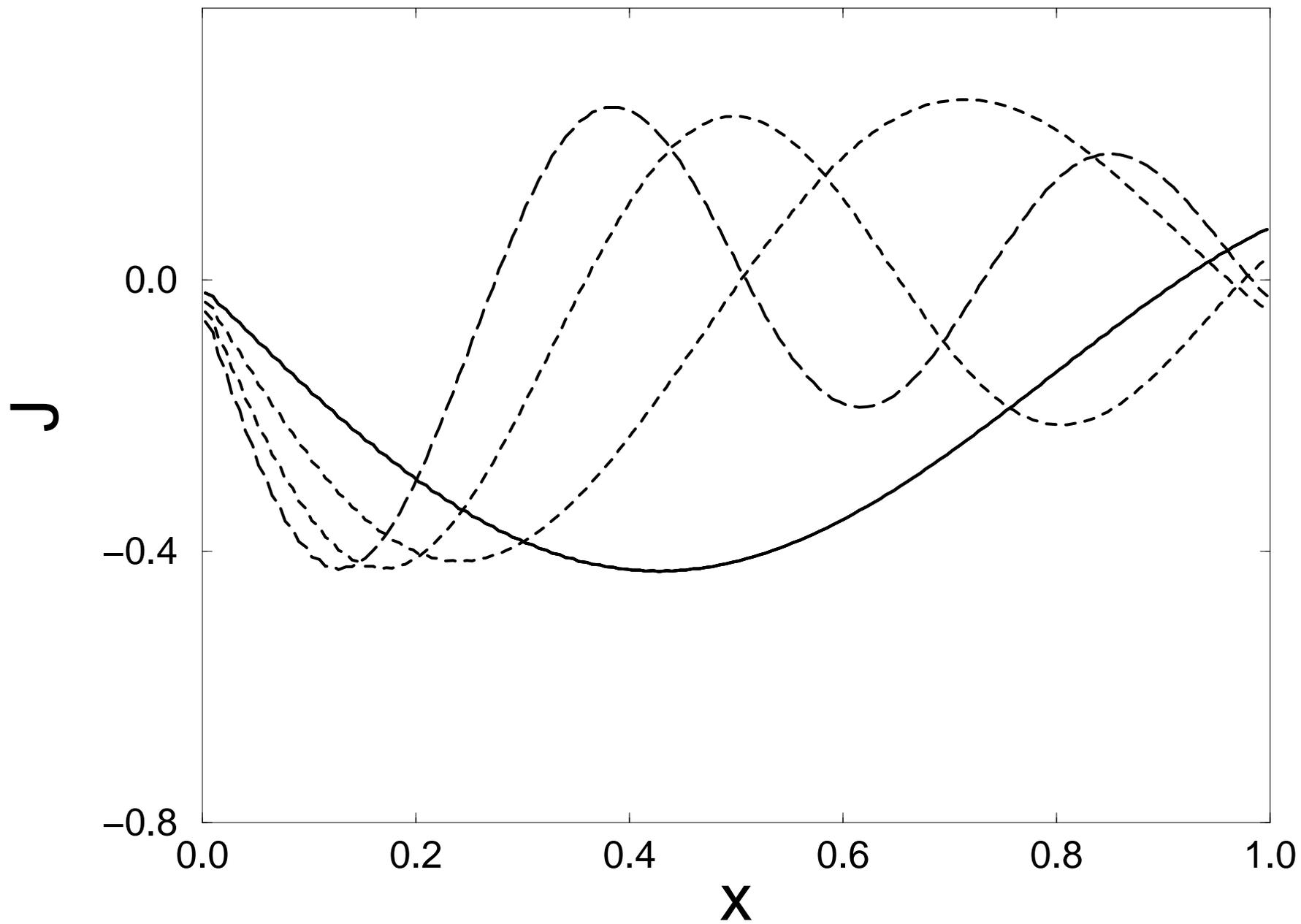
\begin{figure}[t]
\caption{Current distribution in a disk. The
parameters are $qR=1$ and $\alpha=0$
(see equation (\ref{GFDISK})). The four modes
correspond to the four positive eigenvalues
$\lambda_{n} \equiv 1/\mu_{n}$
of equation (\ref{ETAEQ}), with $n=1,2,3,4$ which
also counts the number of radial nodes.\label{Fig5}}
\end{figure}

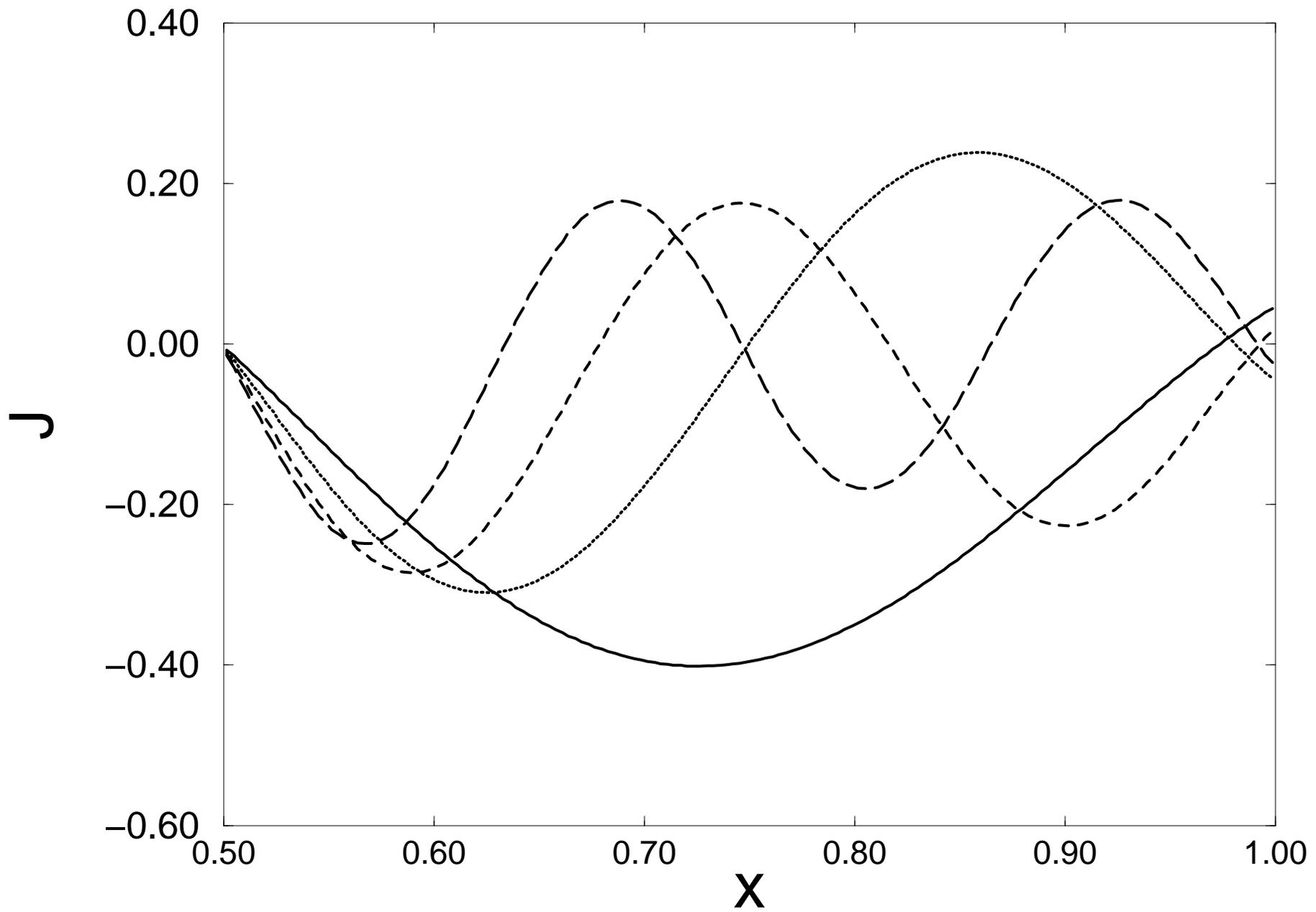
\begin{figure}[t]
\caption{
Current distribution in an annulus\label{Fig6}. The
parameters are $qR_{1}=0.5$, $qR_{2}=1$ and $\alpha=1$,
$\beta=\gamma=\delta=0$ (see equation (\ref{GFANNULUS})).
The four modes are as in figure \ref{Fig5}.}
\end{figure}

We now present our results for the current profile in the
circular geometry. The derivation of the current once the
charge density is given has been explained for the Hall bar
geometry in refs. \cite{KN,SK}. Here we are interested mainly
in the oscillatory pattern. It is not difficult to show that
the tangential current $J_{\theta}(x=qr)$ is given by,
\bea            \label{CURR}
J_{\theta}(x)=a \frac {1} {x}
\int_{qR_{1}}^{x} x' \rho (x') dx' + b,
\enea
where $a$ and $b$ are constants and $R_{1}=0$ for the disk and
finite for the annulus. In the following we display the integral appearing
on the RHS of (\ref{CURR}) since it gives the main characteristics of
the actual current. In figure \ref{Fig5} the current profile for the
disk is shown, for the same configuration as in figure \ref{Fig1}.
The current of the lowest mode is similar
in shape to the current evaluated in the Hall bar
geometry \cite{Macdonald,SK}. The current of other modes is
oscillating, which shows clearly that for these modes, current
is not concentrated solely near the edges. Similar result
occurs also for the annulus geometry, for which the current
of the four modes pertinent to figure \ref{Fig2} is displayed
in figure \ref{Fig6}.
\bigskip

\noindent
One of us (Y.A) would like to thank the Japanese Society
for the Promotion of Science for supporting his visit to
the Institute for Solid State Physics in which this work has
been carried out.

\section*{Appendix A}

\noindent
{\em Solution of the equation for the integer QHE.}\\
In this appendix we present a Green function solution of the
equations derived in Ref.\cite{Macdonald} for the integer quantum
Hall effect in a Hall bar. If the
Hall bar stretches along the $y$ direction between points
$x=L_{1}$ and $x=L_{2}$, the charge density
and the electrostatic potential depend
only on $x$. They are determined by a set of equations,
\bea                            \label{MBR}
&&V(x)=-2 e^{2} \int_{L_{1}} ^{L_{2}} \log|x-x'| \rho(x') dx' \\
&&\rho(x)=\frac {n} {h \omega_{c}} V''(x),
\enea
where $n$ is the Landau level number and $\omega_{c}$ is
the cyclotron frequency. Transforming to dimensionless
coordinates $x \rightarrow qx$
 this set of equations
is equivalent to a single equation
for $\rho(x)$ which, for antisymmetric solutions reads
\bea            \label{MBR1}
- \rho(x)= \mu
\frac{d^{2}} {dx^{2}} \int_{0} ^{q L_{x}}
\log \left| \frac {x-x'} {x+x'}\right| \rho(x') dx',
\enea
where $\mu \equiv (4e^{2} n q)/ (h \omega_{c} )>0$
and $L_{x}=L_{2}=-L_{1}$ is half the
width of the Hall bar. Here $q$ is just a parameter
with  dimensions of inverse length.
In order to apply our Green function method
for solving Eq.(\ref{MBR1}) we add $\rho ''$ on both sides and
use algebraic manipulations as in Section 2. The result, using
the notation of section 2, is
\bea   \label{MBR2}
-G \rho = \mu (1+G) L \rho.
\enea
If we apply the operator $(1+G)^{-1}$ on both sides, we get
\bea         \label{MBR3}
[1-(1+G)^{-1}] \rho = \mu (-L) \rho.
\enea

Equation (\ref{MBR3}) is of the
form (\ref{ev}) with symmetric $A$
and symmetric positive definite $B$ and therefore it has
real eigenvalues $\mu$. It would have been more useful to
solve an equation for the potential $V(x)$ rather than the
density $\rho(x)$, since the later one is known to have
a power singularity near the edges \cite{Macdonald,Thouless,SK}.
Unfortunately we were unable to obtain an equation of
the form (\ref{ev}) with symmetric $A$
and symmetric positive definite $B$.

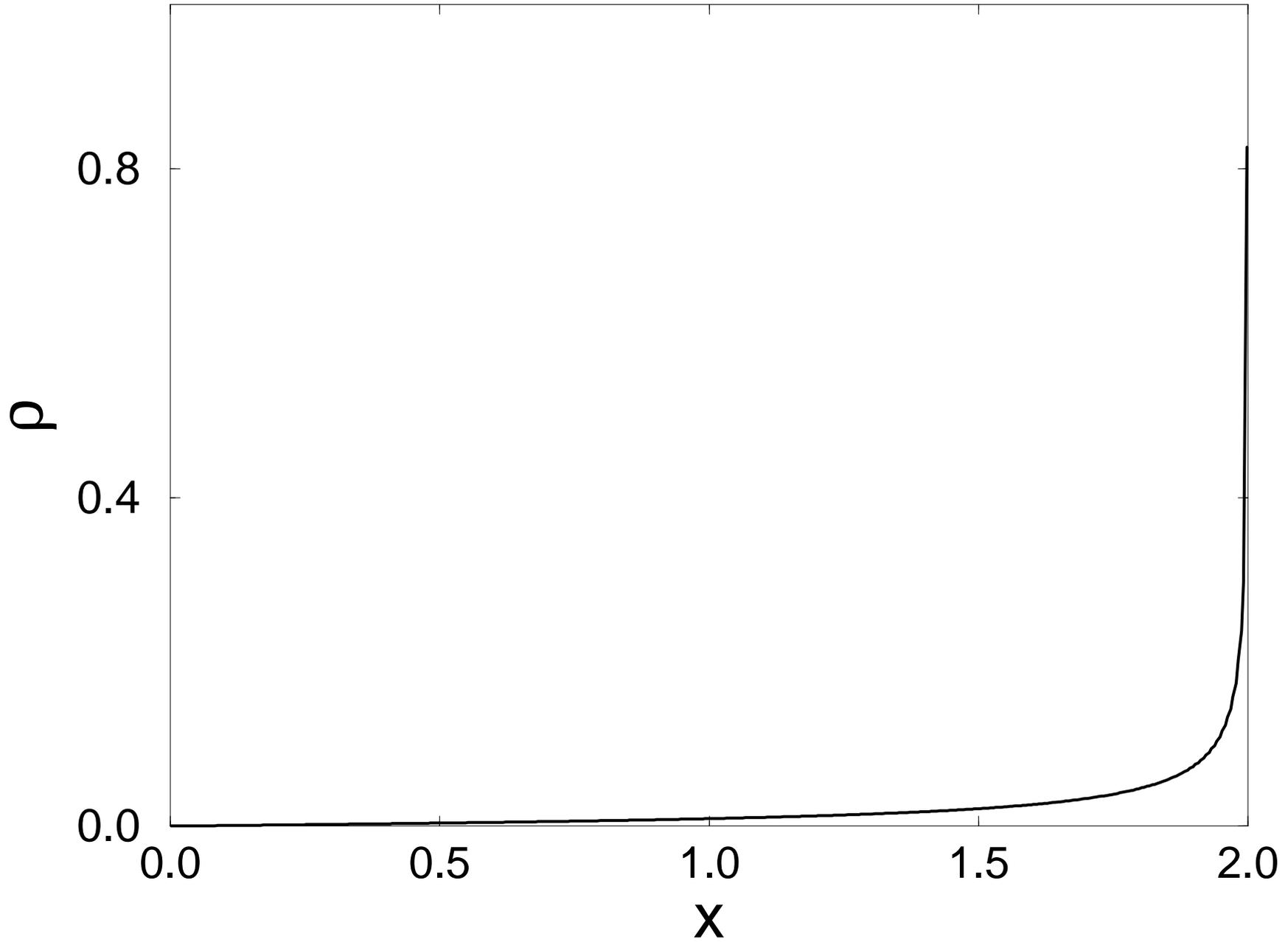
\begin{figure}[t]
\caption{
Charge distribution in a Hall bar for the integer
quantum Hall system\label{Fig3} obtained by solving
equation (\ref{MBR3}). The Hall bar is located between
$x=-2$ and $x=2$ and the charge distribution is antisymmetric.
The parameter $\alpha$ in the Green function (\ref{GF}) is set
equal to zero.}
\end{figure}

\noindent
The charge density in a Hall bar for the integer QHE
is displayed in Fig.~\ref{Fig3}.
For the actual numerical solution we take the Hall bar
to be located between $x=-2$ and $x=+2$ in order to compare
the charge density of the non-interacting system with that
of the interacting system reported below in Appendix B.
As can convincingly be inferred from Fig.~\ref{Fig3}
the charge density in the integer quantum
Hall regime is singular near the edge. This is
of course consistent with earlier results
\cite{Macdonald,Thouless,SK}.

\noindent
\section*{Appendix B}

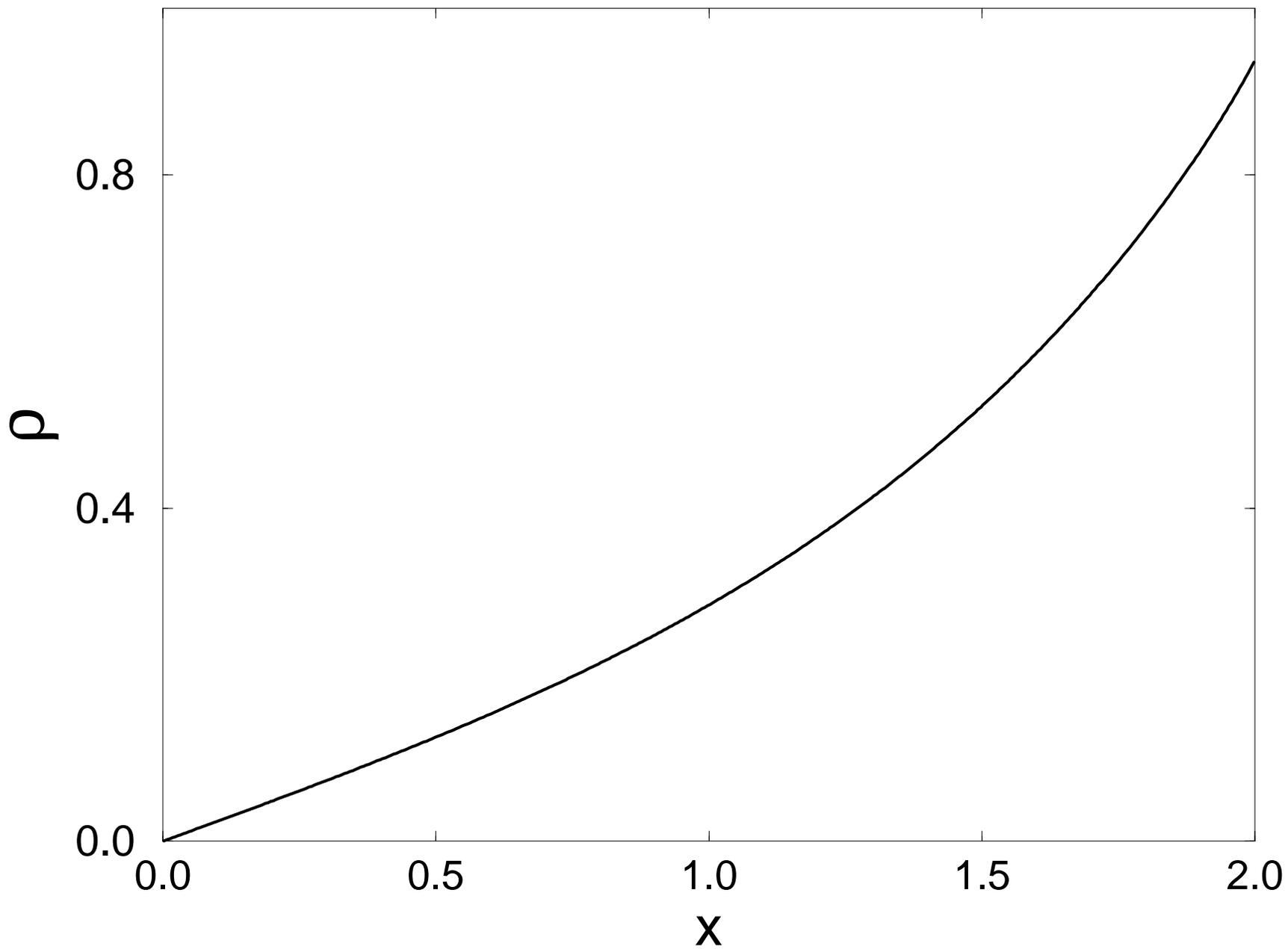
\begin{figure}[t]
\caption{
Charge distribution in a Hall bar for the fractional
quantum Hall system\label{Fig4}
obtained by solving
equation (\ref{HBREAL}). The Hall bar is located between
$x=-2$ and $x=2$ and the charge distribution is antisymmetric.
The parameter $\alpha$ in the Green function (\ref{GF}) is set
equal to zero.}
\end{figure}

\noindent
{\em Charge density profile in the Hall bar geometry
for the fractional QHE.} \\
As was stated already, the charge distribution profile in
the Hall bar geometry for the fractional quantum Hall system
was calculated by an analytic iterative
method \cite{SK}. In this appendix
we present the exact solution obtained within the Green function
technique introduced in Section 2.
We have solved Eq.(\ref{HBREAL}) numerically using mesh
of $N=400$ integration points between $0$ and $2$, replacing
the operators by $N \times N$ matrices. The weak logarithmic
singularity at $x=x'$ is of course integrable. Out of the $N$
eigenvalues only one is found to be positive. This feature
as well as the numerical value of the positive eigenvalue
persist independent on the number $N$ (provided it is
sufficiently large). The parameters $qL$ (upper limit of
integration in Eq.(\ref{INTEQ}) and $\alpha$ (the
free parameter in the Green function (\ref{GF}) should be
chosen in such a way that Eq.(\ref{HBREAL}) has at
least one positive eigenvalue. In all our attempts, we
could not obtain more than a single positive eigenvalue,
a result which is markedly distinct from that obtained
in the circular geometry. For the special choice $qL=2$ and
$\alpha=0$ the charge density in a Hall bar is displayed
in Fig.~\ref{Fig4}. It is reassuring to note that
this solution is extremely close to the one obtained
by iteration methods \cite{SK} or the one obtained by
solving Eq.(\ref{HBREAL}) using a power basis
technique as explained in the next Appendix.

\section*{Appendix C}
\noindent
{\em Solution of the equations in the
Hall bar using a power series expansion.}\\
In this appendix we present an alternative
formulation to study the integro-differential Eq.(\ref{KNHB1})
in the form
\begin{equation}
u(x)-a  u''(x)
=
-
\mu
\int_{-1}^{1} dx' \ln |x-x'| u''(x'), \label{eqfinite}
\end{equation}
with the antisymmetry condition $u(-x)=-u(x)$.
Here, we have used the notation $\rho(x)= u''(x)$.
We can then choose $\{x^{2k+1}|k=0,1,\cdots\}$ as
an expansion basis.
The integration operator on the RHS of (\ref{eqfinite}) is
represented as
\begin{eqnarray}
L \cdot x^{2l+1}&=&-
\int_{-1}^{1} dx' \ln |x-x'| {x'}^{2l+1} \nonumber\\
&=&
\sum_{k=0}^{\infty} \frac{-2}{(2k+1)(2k-2l-1)} x^{2k+1},
\end{eqnarray}
whose matrix representation reads
\begin{eqnarray}
L&=&
\left(
\begin{array}{ccccc}
 2 &  \frac{2}{3} &  \frac{2}{5} & \frac{2}{7} & \cdots \\
-\frac{2}{3} &  \frac{2}{3} & \frac{2}{9} &  \frac{2}{15}&  \\
-\frac{2}{15} & -\frac{2}{5} & \frac{2}{5}&  \frac{2}{15}& \\
-\frac{2}{35} & -\frac{2}{21} & -\frac{2}{7}&  \frac{2}{7}& \\
\vdots &   & & &\ddots
\end{array}
\right),
\end{eqnarray}
and the double derivative is
$$
D^2 {x}^{2l+1} = {d^2 \over dx^2} {x}^{2l+1}=(2 l+1)(2 l) x^{2l-1}.
$$
Thus we have
$$
(1-a D^2) u =\mu   L  D^2 u.
$$
Since the double derivative $D^2$ has rapidly increasing
next-diagonal elements $(2 l+1)(2 l)$,
it seems impossible to simulate the eigenvalue problem (\ref{eqfinite})
numerically with any finite size truncation of $D^2$.
A remedy for this is to integrate twice the
original equation (\ref{eqfinite}).
Let us introduce the double
integration operator $I^2$ by
$$
I^2 x^{2 l+1}=
c_l x + {1 \over (2 l + 3)(2 l+2)} x^{2 l+3},
$$
where $c_l$ is an integration constant. Then all the solutions of
equation
$$
I^2 (I^2 L)^{-1} (I^2 -a) u =\mu u,
$$
also satisfy equation (\ref{eqfinite}). After finding some suitable
set of constants $c_l$, we obtain a positive eigenvalue $\mu$
whose eigenvector is consistent with the one
obtained by the Green function method and by an
analytic argument \cite{SK}.


\end{document}